\documentclass[twocolumn, twocolappendix]{aastex631}

\usepackage{amsmath}
\usepackage{float}
\usepackage{enumitem}
\usepackage{comment}

\newcommand{\ignore}[1]{}

\newcommand*{\rh}[1]{\textcolor{blue}{#1}}
\newcommand*{\cl}[1]{\textcolor{red}{#1}}
\newcommand*{\gm}[1]{\textcolor{green}{#1}}
\newcommand*{\rev}[1]{{#1}}

\renewcommand*{\rh}[1]{#1}
\renewcommand*{\cl}[1]{#1}
\renewcommand*{\gm}[1]{#1}

\shorttitle{H4RG-10 Nonlinearity Correction}
\shortauthors{Hamada et al.}

\graphicspath{{figures/}}

\begin{document}
%\linenumbers

\title{\gm{Nonlinearity in H4RG-10 Near-Infrared Detectors at Elevated Temperatures: Characterization and Data-Driven Correction Method}}

\author[0009-0007-1381-3384]{Ryusei Hamada}
\affiliation{Department of Earth and Space Science, Graduate School of Science, The University of Osaka, 1-1 Machikaneyama-cho, Toyonaka, Osaka 560-0043, Japan}
\correspondingauthor{Ryusei Hamada}
\email{hryusei@iral.ess.sci.osaka-u.ac.jp}

\author[0000-0002-5982-566X]{Gregory Mosby Jr.}
\affiliation{Astrophysics Science Division, NASA Goddard Space Flight Center, 8800 Greenbelt Rd,
Greenbelt, MD 20771, USA}

\author[0000-0001-9818-1513]{Shota Miyazaki}
\affiliation{Institute of Space and Astronautical Science, Japan Aerospace Exploration Agency, 3-1-1 Yoshinodai, Chuo, Sagamihara, Kanagawa 252-5210, Japan}

\author[0000-0002-5843-9433]{Daisuke Suzuki}
\affiliation{Department of Earth and Space Science, Graduate School of Science, The University of Osaka, 1-1 Machikaneyama-cho, Toyonaka, Osaka 560-0043, Japan}

\author[0000-0002-2715-8460]{Alexander Kutyrev}
\affiliation{Astrophysics Science Division, NASA Goddard Space Flight Center, 8800 Greenbelt Rd,
Greenbelt, MD 20771, USA}
\affiliation{Department of Astronomy, University of Maryland, College Park, MD 20742-4111, USA}

\author[0000-0002-3774-1270]{Joseph Durbak}
\affiliation{Astrophysics Science Division, NASA Goddard Space Flight Center, 8800 Greenbelt Rd,
Greenbelt, MD 20771, USA}
\affiliation{Department of Astronomy, University of Maryland, College Park, MD 20742-4111, USA}

\author[0000-0003-4776-8618]{Yuki Hirao}
\affiliation{Institute of Astronomy, Graduate School of Science, The University of Tokyo, 2-21-1 Osawa, Mitaka, Tokyo 181-0015, Japan}

\author[0000-0002-4035-5012]{Takahiro Sumi}
\affiliation{Department of Earth and Space Science, Graduate School of Science, The University of Osaka, 1-1 Machikaneyama-cho, Toyonaka, Osaka 560-0043, Japan}

\begin{abstract}
We report a newly identified nonlinearity in H4RG-10 near-infrared detectors operating under moderately elevated-temperature conditions (\gm{114~K}). This component, \gm{that potentially arises} from illumination-independent defect currents, introduces additional nonlinearity not captured by conventional correction models. To address this issue, we propose a data-driven \gm{nonlinearity} correction (NLC) method that models the nonlinear behavior of both the classical response and the defect currents, using dual-illumination measurements and a dark exposure. Applied to H4RG-10 detectors on the PRIME telescope, the method significantly improves signal linearity, especially for pixels with high defect current, while maintaining comparable performance elsewhere. By selecting the optimal correction model per pixel, reliable NLC is achieved across the full array. This study characterizes \gm{a nonlinearity} intrinsic to H4RG-10 detectors and demonstrates that data-driven post-processing can effectively restore linearity \gm{in the presence of large defect currents.} Although these effects are unlikely to be significant under nominal operating temperatures, the approach may provide a practical calibration framework for future warm-operation scenarios.

\end{abstract}
\section{Introduction} \label{sec:intro}

Over the past two decades, hybrid near-infrared detectors have advanced dramatically, culminating in large-format, high-performance arrays such as the Teledyne H4RG-10\footnote{\rh{The H4RG-10 has a 4096×4096 format with  $10~\mu\mathrm{m}$ pixels and a nominal $2.5~\mu\mathrm{m}$ cutoff wavelength.}}, selected for the \textit{Nancy Grace Roman Space Telescope}.  The HxRG family, H1RG, H2RG (JWST NIRCam), and H4RG employs HgCdTe photodiodes bump-bonded to silicon readout integrated circuits (ROICs) with a source-follower per detector (SFD) architecture. This design enables multiple non-destructive reads and precise measurements under low-background conditions, which are critical for deep infrared imaging. However, it also introduces systematic deviations from ideal linear response, arising from nonlinearities in charge accumulation within the HgCdTe layer, the ROIC voltage response, and downstream digitization effects \citep[e.g.,][]{2020JATIS...6d6001M}. As a result, even under constant illumination, pixel signals deviate from a linear relationship with incident flux. Such effects are typically modeled deterministically and corrected using empirical flat-field–based linearization methods \citep{2017jwst.rept.5167C}, which assume that dark current is negligible or linear in time, a valid assumption only under sufficiently cold and stable operating conditions \gm{provided that the detector material is pristine}.

%Description of H4RG
%Over the past two decades, hybrid near-infrared detectors have advanced markedly, culminating in large-format, high-performance arrays like the Teledyne H4RG-10, selected for the Nancy Grace Roman Space Telescope. These HxRG-family detectors—H1RG (NICMOS), H2RG (Hubble WFC3, JWST), and H4RG—use HgCdTe photodiodes bump-bonded to silicon ROICs with source-follower per detector (SFD) architecture. SFD allows multiple non-destructive reads, enabling high precision under low-background conditions essential for deep infrared imaging. However, this architecture induces systematic deviations from linear response, known as Classical Non-Linearity (CNL), due to nonlinearities in the HgCdTe charge accumulation, ROIC voltage response, and downstream digitization \citep[See][in detail.]{2020JATIS...6d6001M}. Consequently, even under stable illumination, pixel signals deviate from linear scaling with incident flux. Although complex, CNL is typically modeled deterministically and corrected via empirical flat-field-based linearization \citep{2017jwst.rept.5167C}. However, these corrections assume negligible or linear dark current, which is typically valid under cooled and stable conditions.
\begin{figure*}
    \centering
    \includegraphics[width=\textwidth]{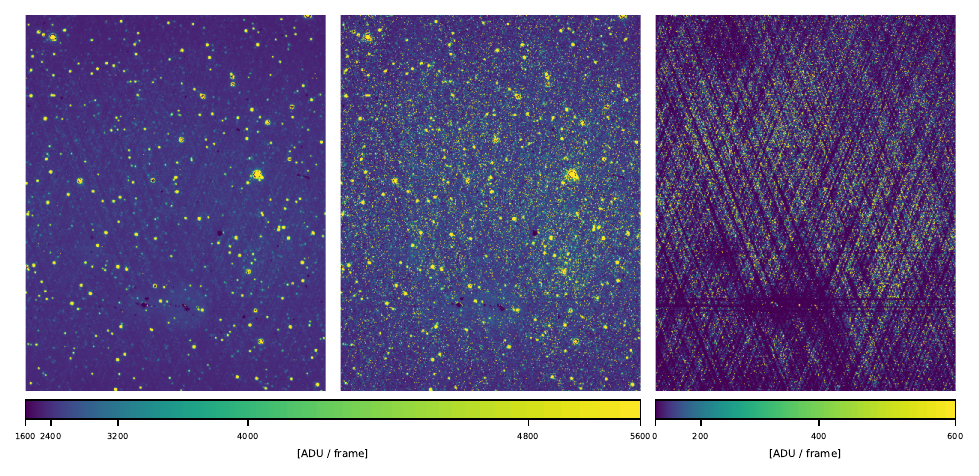}
    \caption{
    {\bf Left}: \rev{A part of an H4RG-10 image observed by the PRIME telescope toward the Galactic bulge at a detector temperature of 114~K, processed with superbias subtraction, reference pixel correction, and ramp fitting, but without applying any nonlinearity correction.} {\bf Middle}: The same dataset processed with superbias subtraction, reference pixel correction, conventional nonlinearity correction (CNLC), \rev{ramp fitting}, and dark subtraction, shows a cross-hatched pattern in certain regions. This pattern is potentially caused by elevated dark current in some pixels, unaccounted for in conventional nonlinearity correction. \rev{The left and middle panels share a common color scale, with a stretch applied using a method similar to the DS9 {\it zscale} + {\it sinh} scaling. {\bf Right}: A dark image obtained at the same detector temperature (114~K) calculated from the slope of ten frames from a dark dataset (exp3), excluding the first frame (frame 0), extracted from the same spatial region as in the left panel.}
    }
    \label{fig:comp_nlc}
\end{figure*}

\begin{figure}[h!]
    \centering
    \includegraphics[width=\columnwidth]{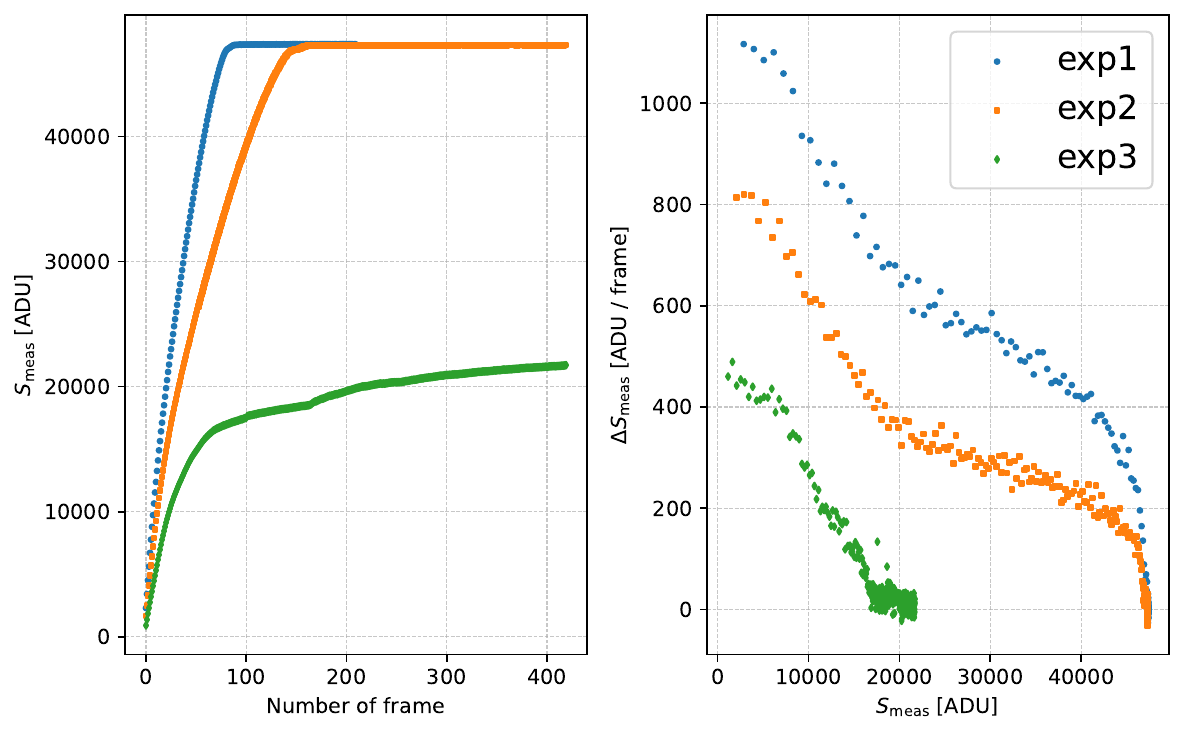}
    \caption{
    {\bf Left}: Measured signal $S_\mathrm{meas}$ in analog-to-digital units (ADU, or DN) as a function of frame number for two illuminated exposures (blue and orange) and one dark exposure (green) taken with a PRIME H4RG-10 detector. {\bf Right}: Corresponding per-frame signal slope \gm{from finite differencing (Eq.~\ref{eq:a.2} )} as a function of $S_\mathrm{meas}$ \gm{(Eq.~\ref{eq:a.1})}. The data corresponds to a representative pixel exhibiting a significant \gm{spurious} dark current. \gm{These behaviors suggest that the signal in this pixel without illumination is not constant but evolves nonlinearly with the accumulation of charge, especially under elevated detector temperatures.}
    }
    \label{fig:picture}
\end{figure}

The limitations of the assumptions of these traditional linearization methods are particularly relevant for the PRIME \citep[PRime-focus Infrared Microlensing Experiment;][]{Kondo+2023, Yama+2023prime, 2025AJ....170..338S} telescope, a 1.8-meter near-infrared survey instrument at the South African Astronomical Observatory designed to monitor the Galactic bulge \gm{for microlensing events}. Its wide-field camera, PRIME-Cam \citep{Kutyrev+2023prime, Durbak+2024prime}, employs four H4RG-10 detectors to cover $\sim1.45\;\mathrm{deg^2}$ and began operations in 2023. However, due to \rh{practical thermal constraints}, stable operation at the nominal \gm{90~K} is unachievable; instead, the detectors are operated at a fixed temperature between \gm{105-125~K}, depending on the ambient conditions. We found that operating in this elevated-temperature regime induces a significant nonlinear \gm{dark} current component, \gm{particularly in regions of \rev{known} defects in the focal plane sensors. Hereafter, we refer to this current as the ``defect current'' though the exact origin of this spurious current is beyond the scope of this paper.} This current introduces additional nonlinearity that invalidates conventional correction schemes.  \gm{Because such behavior is not accounted for in conventional nonlinearity correction (NLC) procedures, applying those corrections can produce or amplify noticeable artifacts in science images, as shown in Figure~\ref{fig:comp_nlc}. These results highlight the necessity of a correction model that incorporates the evolution of the defect signal under elevated detector temperature conditions.}

Figure \ref{fig:picture} illustrates this effect \gm{for a single pixel}. 
The left panel shows the measured signal $S_\mathrm{meas}$ (in ADU or DN) as a function of frame number for two background-illuminated exposures and one dark exposure obtained with the PRIME H4RG-10 detector.\footnote{Here, $S_\mathrm{meas}$ denotes the signal derived from raw frames after superbias subtraction and reference-pixel correction. The background-illuminated frames include thermal emission from the telescope, dome, and filters rather than any artificial illumination.} The right panel plots the per-frame signal slope as a function of $S_\mathrm{meas}$. For pixels exhibiting large dark current, the slope of \gm{the signal in the dark} decreases rapidly with accumulated charge, indicating \gm{strong nonlinearity}.  Motivated by this behavior, we develop a modeling approach that explicitly treats the \rev{defect current component} as a function of accumulated charge.

%Notably, as seen in the right panel of Figure~\ref{fig:picture}, this nonlinear defect current appears to depend not directly on the instantaneous input flux but on the accumulated signal ($S_\mathrm{meas}$) itself. -GM removing this as we don't have sufficient evidence to back this up yet.

This study aims to develop and validate a calibration technique capable of correcting pixel-level nonlinear defect current. We introduce a data-driven NLC model \gm{that captures classical nonlinearity effects that include a nonlinear defect current for individual pixels} and integrate this model into the image correction pipeline. The main contributions of this work are threefold:
\begin{enumerate}
\setlength{\parskip}{0cm} 
\setlength{\itemsep}{0cm} 
    \item[(1)] \gm{Identification of a nonlinearity in H4RG-10 detectors operated at elevated temperatures, potentially arising from defect currents.}
    \item[(2)] Development of a data-driven NLC framework that uses only dual-illumination and dark exposures to model and correct \gm{the nonlinearity.}
    \item[(3)] Introduction of a combined calibration strategy that selects, on a per-pixel basis, between the conventional NLC and the new method, providing a practical pathway \rh{under suboptimal operating conditions}.
\end{enumerate}

The remainder of this paper is organized as follows. Section~\ref{sec:method} presents the mathematical formulation of pixel-level dark and defect currents. Section~\ref{sec:application} describes the modeling framework, data processing procedures, and computational implementation. Section~\ref{sec:result} provides a quantitative evaluation of the proposed method using real PRIME observational data and compares its performance against conventional approaches. Finally, Section~\ref{sec:conclusion} summarizes the findings, discusses current limitations, and outlines future applications of this technique to other instruments affected by similar nonlinear behavior.

\section{Formulation} \label{sec:method}
\subsection{Conventional NLC Method \label{sec:conv_NLC}}

\gm{The conventional nonlinearity correction (CNLC) used for \rh{HxRG sensors in astronomy (e.g., the H1R of Hubble’s WFC3/IR)} and adopted in the JWST pipeline \citep{2011jwst.rept.2163R,2011jwst.rept.2344R,2014wfc..rept...17H,2017jwst.rept.5167C} is summarized as follows.} CNLC assumes that the detector output can be written as a smooth (monotonic) function of \gm{a signal \emph{linear} with time},
\begin{equation}\label{eq:lin_expantion}
    S_{\mathrm{meas},i}=\sum_{k=0}^{K} a_k\,[A t_i]^k,
\end{equation}
with $a_1\!=\!1$. Here $S_{\mathrm{meas},i}$ is the measured ADU in the $i$-th frame (read), $t_i$ is the elapsed time, and $A$ [ADU s$^{-1}$] is a constant count rate that is linear in time over the ramp. In this conventional treatment, $A$ represents the total count rate (illumination plus any approximately constant\gm{, and presumably neglible}, dark current), while $a_0$ accounts for a residual offset not removed by bias subtraction and other terms $a_{k>1}$ encode classical \gm{nonlinearity}.

The aim of CNLC is to construct, from $S_{\mathrm{meas}}$, a linearized signal that grows linearly with time,
\begin{equation}\label{eq:S_lin}
    S_{\mathrm{lin},i}=a_0 + A t_i .
\end{equation}
Operationally, CNLC assumes the existence of a pixel-wise, time-independent mapping $C$ such that
\begin{equation}
    S_{\mathrm{lin}} = C\!\left(S_{\mathrm{meas}}\right),
\end{equation}
where $C$ depends only on the measured ADU and is independent of exposure time or observing conditions. Since $S_{\mathrm{meas}}$ is monotonic in $S_{\mathrm{lin}}$ over the operating range, a local inverse exists and one can write $C$ in the convenient scaling form
\begin{equation}
    C\!\left(S_{\mathrm{meas}}\,|\,b_0,\dots,b_L\right)
    \;=\; S_{\mathrm{meas}}\;\times\;R_{\mathrm{fit}}\!\left(S_{\mathrm{meas}}\right),
\end{equation}
with
\begin{equation}
    R_{\mathrm{fit}}\!\left(S_{\mathrm{meas}}\right)=\sum_{l=0}^{L} b_l\, S_{\mathrm{meas}}^{\,l} .
\end{equation}
The coefficients $\{b_l\}$ are the parameters of the inverse polynomial mapping that converts $S_{\mathrm{meas}}$ to the linearized output $S_{\mathrm{lin}}$. \gm{The $\{b_l\}$} are calibrated per pixel from ramp data by estimating the empirical scale factor $R_{\mathrm{meas}}\!\equiv\! S_{\mathrm{lin}}/S_{\mathrm{meas}}$ and fitting \gm{$R_{\mathrm{meas}}$} as a function of $S_{\mathrm{meas}}$. Once $\{b_l\}$ are determined, any $S_{\mathrm{meas}}$ within the calibrated domain can be corrected via $S_{\mathrm{lin}}=C(S_{\mathrm{meas}})$. It is important to note the implicit physical assumptions behind CNLC: (i) the total count rate is \gm{constant} over a ramp (i.e., any dark current is constant or negligible compared to illumination), (ii) The \gm{nonlinearity} depends only on the instantaneous signal level. When additional illumination-independent, signal-dependent components (e.g., the defect currents) are present, the premise of Eq.~\eqref{eq:lin_expantion} can break down, and extensions of the CNLC model become necessary.

\subsection{Proposed NLC Method including Defect Current} \label{subsec:new_method}
\gm{To complete a nonlinearity correction when the assumptions of the CNLC method breaks down, we derive a new NLC method below that can take into account a spurious, nonlinear dark current source such as a current due to defects.}
\subsubsection{Assumptions}
\gm{We} define the true (input) signal, $S_\mathrm{true}$, as the ADU value that would be measured by an ideal detector free from classical \gm{nonlinearity}, including any defect current component. The following assumptions underlie our model:
\begin{itemize}
\setlength{\parskip}{0cm} 
\setlength{\itemsep}{0cm} 
    \item The measured detector output ($S_\mathrm{meas}$) and the true signal ($S_\mathrm{true}$) have a deterministic one-to-one correspondence.
    \item The measured signal $S_{\mathrm{meas}}$ is approximated by a polynomial expansion of $S_{\mathrm{true}}$, which represents the ideal detector response free from classical \gm{nonlinearity}. $S_{\mathrm{true}}$ consists of an illumination-dependent linear term, $At$, and an illumination-independent defect-current component, $S_{\mathrm{def}}$, \rh{which may be} nonlinear.
    \item The temporal variation of the defect current, $A_\mathrm{def}\;(\equiv \partial S_{\rm def}/\partial t)$, can be expressed as a function of $S_{\rm true}$, and because the NLC assumes a deterministic mapping between input and output signals, this function can equivalently be regarded as a function of $S_{\rm meas}$.
\end{itemize}

Based on these assumptions, the detector output $S_{\mathrm{meas},i}$ corresponding to the $i$-th frame is generalized from Eq.~\eqref{eq:lin_expantion} as follows:
\begin{eqnarray}
    S_{\mathrm{meas},i} = \sum^K_{k=0} a_k \left(S_{\mathrm{true}, i}\right)^k, \label{eq:6}
\end{eqnarray}
where
\begin{eqnarray}
    S_{\mathrm{true},i} = A t_i + \int_{0}^{t_i} A_\mathrm{def}\left(S_{\mathrm{meas}}\right)\;dt. \label{eq:7}
\end{eqnarray}
As in the previous section, we set $a_1 = 1$. Here, $t_i$ denotes the exposure time for the $i$-th frame. The term $A t_i$ represents the component that varies linearly with time, where $A$ \gm{represents the illumination-dependent count rate. Note that the integral term in Eq.~\eqref{eq:7} represents the integral signal over a frame from a potentially time-varying spurious dark current from defects, $A_{\mathrm{def}}$. Though $A_{\mathrm{def}}$ may be time-varying, $S_{\mathrm{meas}}$ is assumed to be monotonically increasing with time, so we parameterize the variation of $A_{\mathrm{def}}$ as a function of $S_{\mathrm{meas}}$}.

\subsubsection{\gm{Retrieving the Constant Illumination-Dependent Count Rate and the Defect Current Components}}
Since both $S_{\mathrm{meas}}$ and $S_{\mathrm{true}}$ can be treated as continuous functions of exposure time $t$, we derive how to estimate $S_{\mathrm{lin}}$ from the function $S_{\mathrm{meas}, A_j}(t)$, where $A = A_j$ represents \gm{the constant, illumination-dependent count rate component}. As a first step, we take the time derivative of $S_{\mathrm{meas}, A_j}(t)$, $\dot{S}_{\mathrm{meas},A_j}(t)\;(\equiv \partial S_{\mathrm{meas},A_j}/\partial t)$:
\begin{eqnarray}
    \dot{S}_{\mathrm{meas},A_j}(t) = \left(A_j + A_\mathrm{def}\left(S_{\mathrm{meas}}\right)\right) \times f_{\mathrm{NL}}\left(S_{\mathrm{meas}}\right), \label{eq:8}
\end{eqnarray}
where
\begin{equation}
    f_{\mathrm{NL}}\left(S_{\mathrm{meas}}\right) = 1 + \sum^{K}_{k=2} k a_k \left(S_{\mathrm{true}}\left(S_\mathrm{meas}\right)\right)^{k-1}. \label{eq:9}
\end{equation}
This equation represents the time derivative of the measured signal, explicitly incorporating the nonlinear contribution from the defect current. \gm{The important point here is that both $A_\mathrm{def}$ and $f_{\mathrm{NL}}$ are uniquely determined once $S_{\mathrm{meas}}$ is given.} Therefore, $\dot{S}_{\mathrm{meas},A_j}$ is also uniquely determined by $S_{\mathrm{meas}}$ and the \gm{illumination-dependent count rate}, $A_j$. We now consider the difference between $\dot{S}_{\mathrm{meas},A_1}$ and $\dot{S}_{\mathrm{meas},A_2}$ corresponding to two distinct illumination levels, $A_1$ and $A_2$:
\begin{eqnarray}
    \left(\Delta \dot{S}_{\mathrm{meas}}\right)_{A_1,A_2} &=& \dot{S}_{\mathrm{meas},A_1}\left(S_\mathrm{meas}\right) - \dot{S}_{\mathrm{meas},A_2}\left(S_\mathrm{meas}\right) \nonumber\\
    &=& (A_1 - A_2) \times f_{\mathrm{NL}}\left(S_{\mathrm{meas}}\right). \label{eq:10}
\end{eqnarray}
In Eq.~\eqref{eq:10}, the defect-current term $A_\mathrm{def}$ cancels out when measurements at two different illumination levels are compared. Moreover, since $f_{\mathrm{NL}}(S_{\mathrm{meas}})$ asymptotically approaches unity as $S_{\mathrm{meas}}$ decreases, the difference $(A_1 - A_2)$ can be estimated from the $\Delta \dot{S}_{\mathrm{meas}}$ samples in the low-$S_{\mathrm{meas}}$ region.

Once $(A_1 - A_2)$ is determined, the following relation holds for a given value of $S_{\mathrm{meas}}$, derived from Eqs.~\eqref{eq:8} and \eqref{eq:10}:
\begin{eqnarray}
    A_1 + A_\mathrm{def}(S_{\mathrm{meas}}) &=& (A_1 - A_2) \cdot \left(1 - S_\mathrm{ratio}\right)^{-1}, \label{eq:11}
\end{eqnarray}
where
\begin{eqnarray}
    S_\mathrm{ratio}(S_{\mathrm{meas}}) \equiv \left.\frac{\dot{S}_{\mathrm{meas},A_2}}{\dot{S}_{\mathrm{meas},A_1}}\right|_{S = S_\mathrm{meas}}. \label{eq:S_ratio}
\end{eqnarray}
Therefore, the sum of the \gm{illumination-dependent, constant count rate component} and the defect-current accumulation rate at a given $S_\mathrm{meas}$ (i.e., $A_1 + A_\mathrm{def}(S_{\mathrm{meas}})$) can be determined. According to Eq.~\eqref{eq:7}, integrating this quantity over time yields $S_{\mathrm{true}}$. \gm{Furthermore, adding the bias term gives the detector output that would be obtained in the absence of classical \gm{nonlinearity}, $S_{\mathrm{lin+def}}$, corresponding to Eq.~\eqref{eq:S_lin}}:
\begin{eqnarray} 
    S_{\mathrm{lin+def}, A_1} &=& a_0 + S_{\mathrm{true}, A_1} \nonumber\\ 
    &=&a_0 + \int_{0}^{t_\mathrm{exp}} \left( A_1 + A_\mathrm{def}(S_{\mathrm{meas}}) \right)\,dt \nonumber \\ 
    &=& a_0 + (A_1 - A_2) \int_{0}^{t_\mathrm{exp}} \left(1 - S_\mathrm{ratio}\right)^{-1}dt .\;\;\;\;\;\; \label{eq:13} 
\end{eqnarray}
Note that, depending on the form of the \gm{defect-current,} $A_\mathrm{def}$, $S_{\mathrm{lin+def}}$ is not necessarily linear. \ignore{\cl{It is also assumed that $S_{\mathrm{lin+def}}$ can be deterministically derived from $S_\mathrm{meas}$.} \footnote{In practice, to evaluate $S_{\mathrm{lin+def}}$, experimental data such as $(A_1 - A_2)$, $\dot{S}_{\mathrm{meas},A_1}$, and $\dot{S}_{\mathrm{meas},A_2}$ obtained from measurements at different illumination levels ($A_1$ and $A_2$) are required.}}

Building on this, we determine, for each pixel, the transform function $R:S_\mathrm{meas} \rightarrow S_\mathrm{lin+def}$. From the definition in Eq.~\eqref{eq:8}, $S_\mathrm{meas}$ is obtained by integrating $\dot{S}_\mathrm{meas}$ over time. Here, we consider the case of a light source that produces a count rate of $A_1$, yielding:
\begin{eqnarray}
    S_{\mathrm{meas},A_1} = a_0 + \int_0^{t_\mathrm{exp}} \dot{S}_{\mathrm{meas},A_1}\,dt. \label{eq:14}
\end{eqnarray}
The corresponding ratio $R_\mathrm{meas}$ for a given $S_\mathrm{meas}$ is then expressed as:
\begin{align}
    R_\mathrm{meas} &\equiv \frac{S_{\mathrm{lin+def}}}{S_{\mathrm{meas}}} \nonumber\\
    &= \frac{a_0 + (A_1 - A_2) \int_{0}^{t_\mathrm{exp}} \!\left(1 - S_\mathrm{ratio}\right)^{-1} dt}{a_0 + \int_0^{t_\mathrm{exp}} \dot{S}_{\mathrm{meas},A_1}\,dt} \notag \\
    &= 1 + \frac{\int_{0}^{t_\mathrm{exp}} \!\left[(A_1 - A_2)(1 - S_\mathrm{ratio})^{-1} - \dot{S}_{\mathrm{meas},A_1}\right] dt}{S_{\mathrm{meas},A_1}}. \label{eq:Rmeas}
\end{align}
By fitting \( R_\mathrm{meas} \) as a function of \( S_\mathrm{meas} \), we obtain the empirical correction function \( R_\mathrm{fit}:S_\mathrm{meas} \rightarrow S_\mathrm{lin+def} \).

\subsubsection{Removal of the Defect-Current Contribution}
Although $R_\mathrm{fit}(S_\mathrm{meas})$ removes the classical \gm{nonlinearity}, $S_{\mathrm{lin+def}}$ may still exhibit residual nonlinearity when a \gm{non-negligible amount of varying defect current is present}. Therefore, to obtain a purely linear response, the accumulated defect-current contribution must be removed.

We define the accumulated defect-current component as
\begin{equation}
    S_\mathrm{def} \equiv \int_0^{t_\mathrm{exp}} A_\mathrm{def}(S_\mathrm{meas})\;dt, \label{eq:Sdef}
\end{equation}
where $A_\mathrm{def}(S_\mathrm{meas})$ represents the instantaneous defect-current rate, which may itself be nonlinear in $S_\mathrm{meas}$. 
Using $R_\mathrm{fit}$ derived from the illuminated datasets, we can reconstruct the detector response under dark conditions ($A\approx0$), where only \gm{the spurious dark current (e.g., defect current) contributes}. Applying $R_\mathrm{fit}$ to the dark sequence $\{S_\mathrm{meas, A\approx0}\}$ thus yields the accumulated defect-induced \gm{signal}, $a_0 + S_\mathrm{def}$. Taking the time derivative gives the instantaneous defect-current rate,
\begin{equation}
    A_\mathrm{def}(S_{\mathrm{meas}, A\approx0}) = \frac{\partial}{\partial t} \left[ S_{\mathrm{meas}, A\approx0} \times R_{\mathrm{fit}}\left(S_{\mathrm{meas}, A\approx0}\right) \right], \label{eq:Adef}
\end{equation}
which links the dark-frame slope directly to the nonlinearity model.

The derived \( A_\mathrm{def}(S_\mathrm{meas}) \) is approximated by a polynomial function, yielding \( A_{\mathrm{def,fit}}(S_\mathrm{meas}) \). Integrating this function over time provides the accumulated defect current, \( S_\mathrm{def} \), contained in \( S_{\mathrm{lin+def}} \), and the purely linear signal is then obtained as
\begin{eqnarray}
    S_\mathrm{lin} &=& S_{\mathrm{lin+def}} - S_\mathrm{def} \notag\\
    &=& S_\mathrm{meas} \times R_\mathrm{fit}(S_\mathrm{meas})
    - \int_0^{t_\mathrm{exp}} A_{\mathrm{def,fit}}\,dt. \label{eq:18}
\end{eqnarray}
This formulation enables recovery of the ideal linear signal under continuous conditions. In practice, however, observational data are discretely sampled, and additional considerations for numerical implementation are discussed in Appendix~\ref{ap:A}.

\section{Application} \label{sec:application}

\begin{figure}[t!]
    \centering
    \includegraphics[scale=0.55]{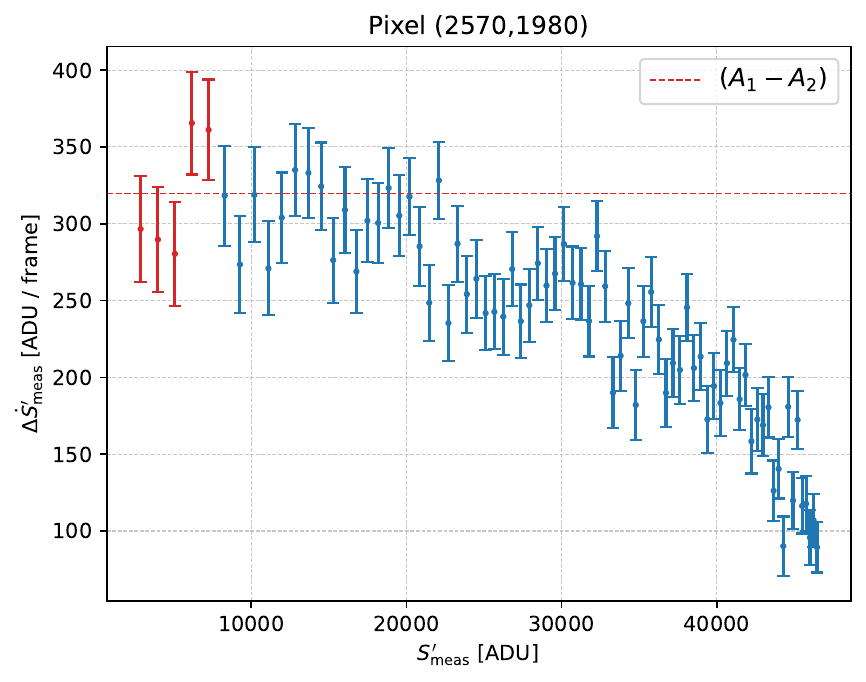}
    \caption{
    Per-pixel differences in measured slopes, $\Delta\dot{S}_\mathrm{meas}$, plotted as a function of $S_\mathrm{meas}$ for a representative pixel exhibiting large defect current. The weighted mean of the five points at lower $S_\mathrm{meas}$ (shown in red) is used to estimate the factor of $(A_1 - A_2)$ in Eq.~\eqref{eq:10}.
    }
    \label{fig:delta_Sdot}
\end{figure}

\begin{figure}[t!]
    \centering
    \includegraphics[scale=0.55]{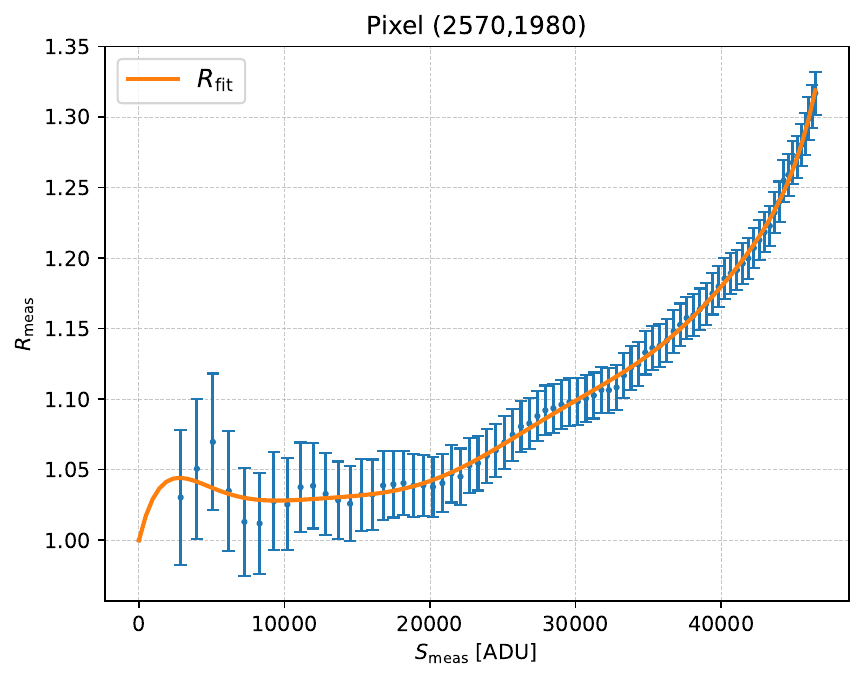}
    \caption{
    The blue \gm{points with error bars} represent the correction factor $R_\mathrm{meas}$ defined in Eq.~\eqref{eq:Rmeas}, plotted as a function of the measured ADU value $S_\mathrm{meas}$. The orange line shows the fitted function $R_\mathrm{fit}: S_\mathrm{meas} \rightarrow S_\mathrm{lin+def}$ with polynomial expansion of $S_\mathrm{meas}$.
    }
    \label{fig:Rmeas}
\end{figure}

\gm{In this section, we demonstrate the application of the newly developed NLC framework to H4RG-10 detector data obtained with the PRIME-Cam.} We first describe the calibration datasets and observing conditions used in the analysis (Section \ref{sec:3.1}). We then construct the NLC model based on the proposed method, illustrating the procedure for representative pixels exhibiting both high and low dark currents (Section \ref{sec:3.2}). \gm{ We present the quantitative evaluation of the correction performance and the resulting improvements in signal linearity in Section~\ref{sec:result}}.

%In this section, we first describe the dataset employed in our analysis. We then illustrate the construction of the non-linearity model based on the newly developed method, focusing on representative cases: pixels exhibiting high dark currents and those with minimal dark currents. Finally, we assess the linearity of the corrected signals across all pixels, comparing the performance of the conventional method with that of the proposed approach.

\subsection{Data Overview}\label{sec:3.1}

\begin{deluxetable}{l|ccc}[t!]
\tablenum{1}
\tablecaption{Summary of the calibration datasets. \label{tab:tab1}}
\tablehead{
\colhead{Conditions} & \colhead{exp1} & \colhead{exp2} & \colhead{exp3}
}
\startdata
%Test Date & \multicolumn{3}{c}{2025 Mar 20} \\
Detector & \multicolumn{3}{c}{H4RG-10 (chip 4)} \\
Detector Temperature [K] & \multicolumn{3}{c}{114.00 $\pm$ 0.01} \\
Filter & $H$-band & $J$-band & Dark \\
Pixel Index $(x, y)$ & \multicolumn{3}{c}{(2570, 1980)} \\
Frame Interval [s] & \multicolumn{3}{c}{2.86} \\
Total Exposure [s] & 600 & 1200 & 1200 \\
\enddata
\end{deluxetable}

\rh{Three calibration datasets with different illumination levels --- hereafter referred to as exp1, exp2, and exp3 --- were acquired with the PRIME-Cam instrument on March~20, 2025, using the H4RG-10 detector (chip~4) maintained at a stable temperature of 114~K.}
\begin{comment}
    Three calibration datasets with different illumination levels, hereafter referred to as exp1, exp2, and exp3, were obtained using the PRIME-Cam instrument. \rh{The data were taken on March 20, 2025,} with the H4RG-10 detector (chip 4) operated at a stable temperature of \gm{114~K}.
\end{comment}
The $H$- and $J$-band datasets (exp1 and exp2) were taken under thermally stable conditions inside the dome, while exp3 was obtained using a dark filter (solid aluminum blank) in the \gm{PRIME-Cam} to block incident light. \rh{The key parameters of the calibration datasets are summarized in Table~\ref{tab:tab1}.} 

During the $H$- and $J$-band measurements, the detector was illuminated by thermal emission transmitted through the band-pass filters and the camera window. \rh{Each exposure was preceded by five detector resets, and ten short (11.44 sec) frames were obtained before and after each long integration. Including the resets, these sequences provide at least $\sim$260 sec of data immediately before and after each long integration, allowing persistence decay and background stability to be monitored.} We confirmed that the illumination variations for all datasets are negligible compared with the level of the classical nonlinearity.
%The frame-to-frame illumination variations were $1.2\times10^{-6}$ for exp1 and $1.3\times10^{-6}$ for exp2, negligible compared with the signal deviations introduced by the classical nonlinearity.

All datasets were calibrated following procedures: superbias subtraction, reference-pixel correction (RPC), residual-offset alignment\footnote{\rh{For exp1 and exp2, we perform a linear fit to the 5 frames following the initial readout, and for exp3, to the 10 frames following the initial readout, to align the offsets between exposures that could not be fully removed by superbias subtraction and RPC. These offsets are typically on the order of a few ADU.}}, and saturation masking. The resulting calibrated sequences, denoted as $\{S_{\mathrm{meas},A_1}\}$, $\{S_{\mathrm{meas},A_2}\}$, and $\{S_{\mathrm{meas},A_3}\}$ for exp1, exp2, and exp3, respectively, form the basis of the subsequent modeling (Section~\ref{sec:3.2}).

\subsection{Model construction using the new method} \label{sec:3.2}

The new NLC model developed in Section \ref{subsec:new_method} explicitly includes a nonlinear defect-current term in addition to the classical nonlinearity. This extension captures \gm{the non-constant} behavior of defect currents, but the \rh{$A_{\mathrm{def}}$ fit} becomes unstable when the defect current is weak and the signal variation is small. For such pixels, we apply a \rh{scheme similar to CNLC as described in Section~\ref{sec:3.2.2}. Pixels} with strong defect currents where the fit is stable are modeled directly following Section \ref{subsec:new_method}. %Pixels are classified as ``high'' or ``low'' defect current based on the signal overlap between ${S_{\mathrm{meas},A_1}}$ and ${S_{\mathrm{meas},A_3}}$. 
The following subsections summarize the modeling framework: computation of $R_{\mathrm{meas}}$, fitting for high-defect-current pixels, and the additional steps for low-defect-current pixels. 

\subsubsection{Modeling of $R_{\mathrm{meas}}$\label{sec:3.2.1}} 

\begin{figure}[t!]
    \centering
    \includegraphics[scale=0.55]{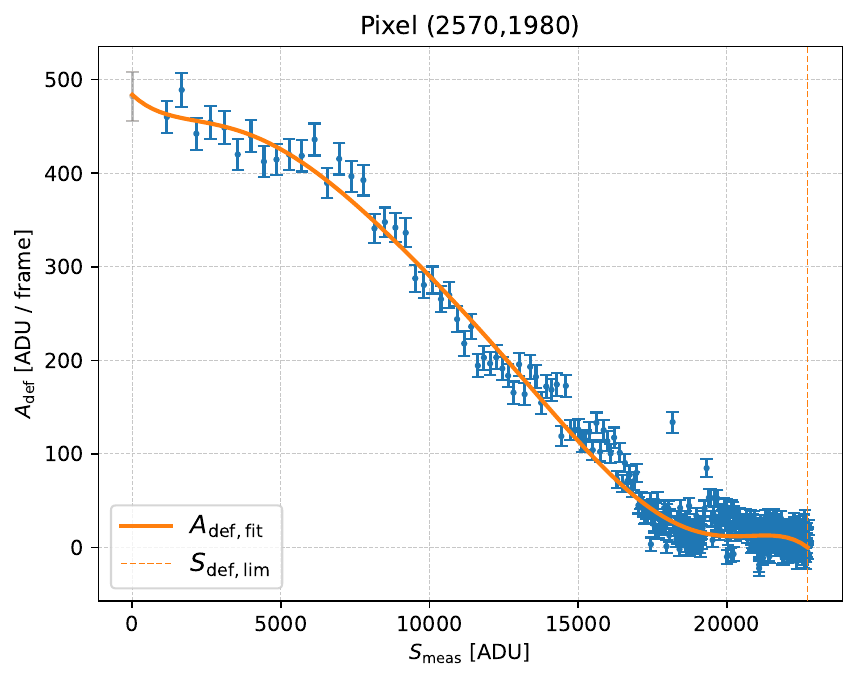}
    \caption{
    Estimated defect current $A_{\mathrm{def}}$ (blue points) and its polynomial fit $A_{\mathrm{def,fit}}$ (orange line) at a representative pixel with strong defect current. The dashed line indicates the \gm{threshold, $S_{\mathrm{def,lim}}$,} beyond which the \gm{$A_{\mathrm{def}}$ fit} is held constant.}
    \label{fig:Adef_fit}
\end{figure}

As shown in Section~\ref{subsec:new_method}, the new NLC method enables the calculation of $R_{\mathrm{meas}} \equiv S_{\mathrm{lin+def}}/S_{\mathrm{meas}}$, which represents the ratio between the idealized signal (including the defect current) and the measured signal. This quantity can be derived directly from two illumination datasets, $\{S_{\mathrm{meas},A_1}\}$ and $\{S_{\mathrm{meas},A_2}\}$, without assuming an explicit analytical form for the defect current $A_{\mathrm{def}}$, as long as $A_{\mathrm{def}}$ can be expressed as a function of $S_{\mathrm{meas}}$.

The key relation of Eq.~\eqref{eq:10} is obtained by taking the difference in the measured signal slopes \gm{from finite differencing (Eq.~\ref{eq:a.2})} between two illumination levels. Figure~\ref{fig:delta_Sdot} shows the observed $\Delta \dot{S}_{\mathrm{meas}}$ as a function of $S_{\mathrm{meas}}$, demonstrating the nonlinear behavior of the detector output. The factor $(A_1 - A_2)$ is estimated \rh{using a weighted average of multiple low-signal data points,} where the classical nonlinearity is negligible and thus $f_{\mathrm{NL}}$ becomes unity. \gm{The weights are determined by the estimated variance of each point}. Using  \rh{the factor $(A_1 - A_2)$} and the ratio of the two input illumination levels ($S_{\mathrm{ratio}}$), $R_{\mathrm{meas}}$ is computed for each $S_{\mathrm{meas}}$ value by Eq.~\eqref{eq:Rmeas}. The derived $R_{\mathrm{meas}}$ values are then fitted with a low-order polynomial of $S_{\mathrm{meas}}$ to obtain a smooth mapping function $R_\mathrm{fit}: S_\mathrm{meas} \rightarrow S_\mathrm{lin+def}$, shown in Figure~\ref{fig:Rmeas}.

\subsubsection{Modeling of the nonlinear defect current \label{sec:3.2.2}}

The defect current $A_{\mathrm{def}}$ is derived following Eq.~\eqref{eq:Adef} and fitted with a polynomial function $A_{\mathrm{def, fit}}(S_{\mathrm{meas}})$ that includes an intercept determined from the low-signal data. %(Appendix~\ref{ap:C} details the fitting constraints and justification) (koko mo tekitou!!!). 
The resulting $A_{\mathrm{def}}$ profile and its fitted curve are shown in Figure~\ref{fig:Adef_fit}. Empirically, pixels with strong defect currents show large $A_{\mathrm{def}}$ values at low $S_{\mathrm{meas}}$ that gradually flatten toward higher signals, remaining only a few tens ADU per frame at large $S_{\mathrm{meas}}$. This trend is consistently observed in our measurements, and it seems that the nonlinear defect current \gm{saturates at a given $S_\mathrm{meas}$}.

Because the dark-exposure data (exp3) do not span the entire dynamic range, the fit to $A_{\mathrm{def}}$ becomes unreliable at large $S_{\mathrm{meas}}$. To avoid unreliable extrapolation, we define $S_{\mathrm{def,lim}}$ as the highest signal level where $A_{\mathrm{def,fit}}$ remains positive, which is shown as a vertical dotted line in Figure~\ref{fig:Adef_fit}. For $S_{\mathrm{meas}} > S_{\mathrm{def, lim}}$, $A_{\mathrm{def,fit}}$ is fixed to $A_{\mathrm{def,fit}}(S_{\mathrm{def,lim}})$. This truncation provides a smooth transition between the physically constrained regime and the high-signal limit where defect currents are almost negligible.

For pixels with low defect currents\footnote{\rh{We consider a pixel to have low defect current if all signals measured in exp3 remain below the exp1 signal \gm{$\sim$45~s} into the exposure.}}
, the new NLC model often yields poorer linearity than the conventional approach. This is likely because its higher flexibility, intended to capture nonlinear defect behavior, becomes excessive for nearly linear pixels, causing overfitting and numerical instability in $R_\mathrm{fit}$. To mitigate \gm{overfitting}, we apply an iterative correction consisting of three steps: (i) an initial $R_\mathrm{fit}$ is applied to correct $S_\mathrm{meas}$; (ii) the corrected signal is smoothed with a first-order fit to suppress noise fluctuations; and (iii) $R_\mathrm{fit}$ is recalculated from the smoothed signal to obtain a more stable mapping. For pixels with extremely weak defect currents, $A_\mathrm{def,fit}$ is treated as a constant, obtained from the slope of a first-order fit to the first ten data points of $\{S_{\mathrm{meas},A_3}\}$.
%For pixels with extremely weak defect currents, $A_\mathrm{def,fit}$ is treated as a constant. In this case, a first-order polynomial is fitted to the first five data points of $\{S_{\mathrm{meas}, A\approx0}\}$, and the slope of this fit is adopted as the constant value of $A_\mathrm{def, fit}$.

\section{Result and Evaluation} \label{sec:result}

\subsection{Linearity evaluation at representative pixels}

\begin{figure}[t!]
    \centering
    \includegraphics[scale=0.55]{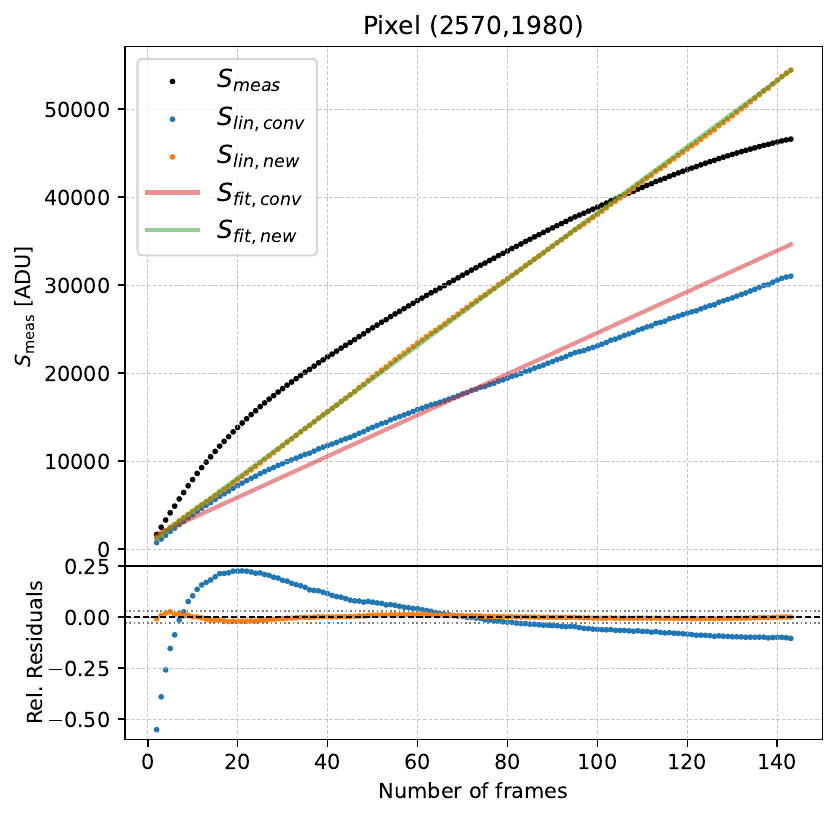}
    \caption{
    Relative residuals from linear fits after applying each correction method \gm{for a single pixel}. \rh{Black, blue, and orange points represent} the uncorrected frame measurements $S_{\rm meas}$, the corrected measurements with the conventional method $S_{\rm lin, conv}$ and the new method $S_{\rm lin, new}$, respectively. The pink and light-green lines represent the linear-fit results for $S_{\rm lin, conv}$ and $S_{\rm lin, new}$, respectively.
    The bottom panel shows the deviations of the corrected measurements from the linear-fit results normalized to the corrected signals.
    }
    \label{fig:frame_vs_Scorr}
\end{figure}

Using the correction models constructed in Section~\ref{sec:3.2}, we now evaluate their performance by applying both the conventional and new NLC methods to the measured datasets. In particular, we examine how effectively the new method restores signal linearity compared to the conventional approach, focusing on representative pixels with different defect-current levels.

We evaluate the effectiveness of the nonlinearity correction by applying both the conventional and new models to $\{S_{\mathrm{meas}, A2}\}$.\footnote{Ideally, linearity should be evaluated using an independent dataset not used in model construction, but $S_{\mathrm{meas}, A_2}$ is \rh{adopted in the current absence of an additional, appropriately stable data set.}} For the new method, the corrected signal is obtained as $S_{\rm lin,new} = S_{\mathrm{meas}}\cdot R_{\mathrm{fit}} - \int A_{\mathrm{def,fit}}\,dt$, whereas for the conventional method the defect-current term is not subtracted and the corrected signal $S_\mathrm{lin, conv}$ is obtained as described in Section~\ref{sec:conv_NLC}. Assuming uncorrelated frame-to-frame variations, a first-order polynomial is fitted to each corrected signal. As a quantitative metric of linearity, we adopt the mean absolute value of the relative residuals between the fitted linear model and the actual $S_{{\rm lin}}$ values. This is denoted as
\begin{equation}
    L_{\rm rel} = \frac{1}{N} \sum^{N}_{i=1} \left| \frac{S_{{\rm lin}, i} - S_{{\rm fit}, i}}{S_{{\rm fit}, i}} \right|,\notag
\end{equation}
where $S_{{\rm fit}, i}$ denotes the fitted value from the first-order polynomial model corresponding to each $S_{{\rm lin}, i}$ and $N$ is the total number of frames.

Figure~\ref{fig:frame_vs_Scorr} shows the result for a representative pixel with high defect current. Both methods yield broadly linear trends, but the conventional method exhibits systematic deviations exceeding $\pm10\%$ at low and high signal levels. On the other hand, the new method suppresses these deviations, keeping residuals within $\pm3\%$ across the sequence. The corresponding linearity metrics, $(L_{\mathrm{rel,conv}}, L_{\mathrm{rel,new}}) = (0.094, 0.0066)$, confirm the improved linearity of the new method. For a low-defect-current pixel at $(x, y) = (2595, 1300)$, the metrics $(0.0025, 0.0033)$ indicate comparable performance between the two methods.

\rh{In addition, $L_{\rm rel}$ is also used to determine the \gm{optimal} polynomial orders required for fitting 
$R_{\rm fit}$, $A_{{\rm def}, {\rm fit}}$, and the conventional method’s $S_{\rm meas}$ 
(denoted as $d_R$, $d_{A_{\rm def}}$, and $d_S$, respectively). 
Specifically, 500 pixels are randomly selected within the same detector, and for all 
combinations of polynomial orders from 4 to 10, we compute $L_{{\rm rel}, {\rm comb}}
\equiv \min\left( L_{{\rm rel}, {\rm conv}},\, L_{{\rm rel}, {\rm new}} \right)$. While we identify the combination that yields the smallest average 
$L_{{\rm rel},{\rm comb}}$ over the 500 pixels, we also confirm that nearby orders 
provide similarly good linearity, and we choose the orders from this stable region. As a result, in this analysis we adopt 
$(d_R, d_{A_{\rm def}}, d_S) = (9, 7, 8)$.}

\subsection{Statistical performance across the detector}

\begin{figure}[t!]
    \centering
    \includegraphics[width=\linewidth]{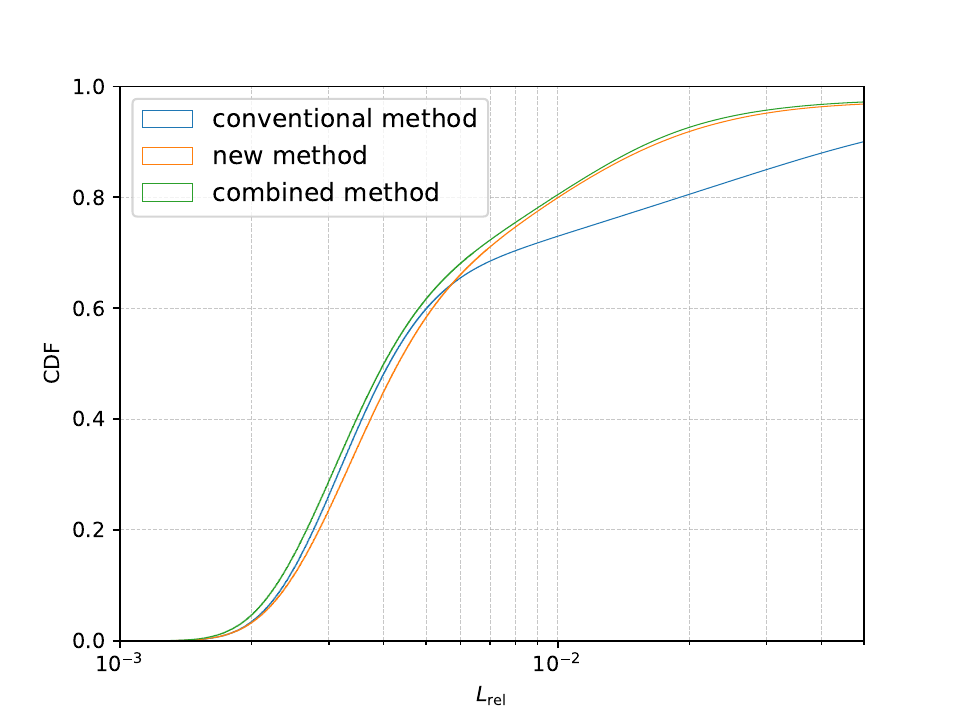}
    \caption{
    Cumulative distribution functions of $L_{\mathrm{rel}}$ for different correction methods, applied across $4088 \times 4088$ pixels of an H4RG-10 chip. \rh{The combined method adopts the method --- either conventional or new --- that gives the lower $L_{\rm rel}$ value on a per-pixel basis.}
    }
    \label{fig:CDF_Lrel}
\end{figure}

We extend the analysis to all $4088\times4088$ pixels of the H4RG-10 detector of PRIME-Cam (chip 4), excluding the reference pixels. Both of the NLC models were successfully constructed for 97.9\% of pixels, for which both $L_{\rm rel}$ and $\chi^2_{S_{\rm ratio}}$ (introduced later) were computed. Figure~\ref{fig:CDF_Lrel} shows the cumulative distribution of $L_{\rm rel}$ for different correction schemes. The resulting curves reveal that the optimal correction method varies across pixels: \gm{the conventional approach performs slightly better for highly linear pixels ($L_{\rm rel}\lesssim0.005$), whereas the new method yields substantially lower $L_{\rm rel}$ values for less linear pixels ($L_{\rm rel}\gtrsim0.005$), reflecting differences in local defect-current properties.} To account for this, we adopt the ``combined method'' on a per-pixel basis, where we choose whichever model yields the lower $L_{\rm rel}$. The fractions of pixels satisfying $L_\mathrm{rel}<0.03$ are 85.0\%, 95.2\%, and 95.7\% for the conventional, new, and combined methods, respectively. These results demonstrate that the new method is broadly applicable and that the pixel-wise selection strategy provides a simple yet effective means to maximize linearity across the array without additional modeling cost.

\subsection{Model selection under the breakdown of conventional assumptions}
\begin{figure}[t!]
    \centering
    \includegraphics[width=\linewidth]{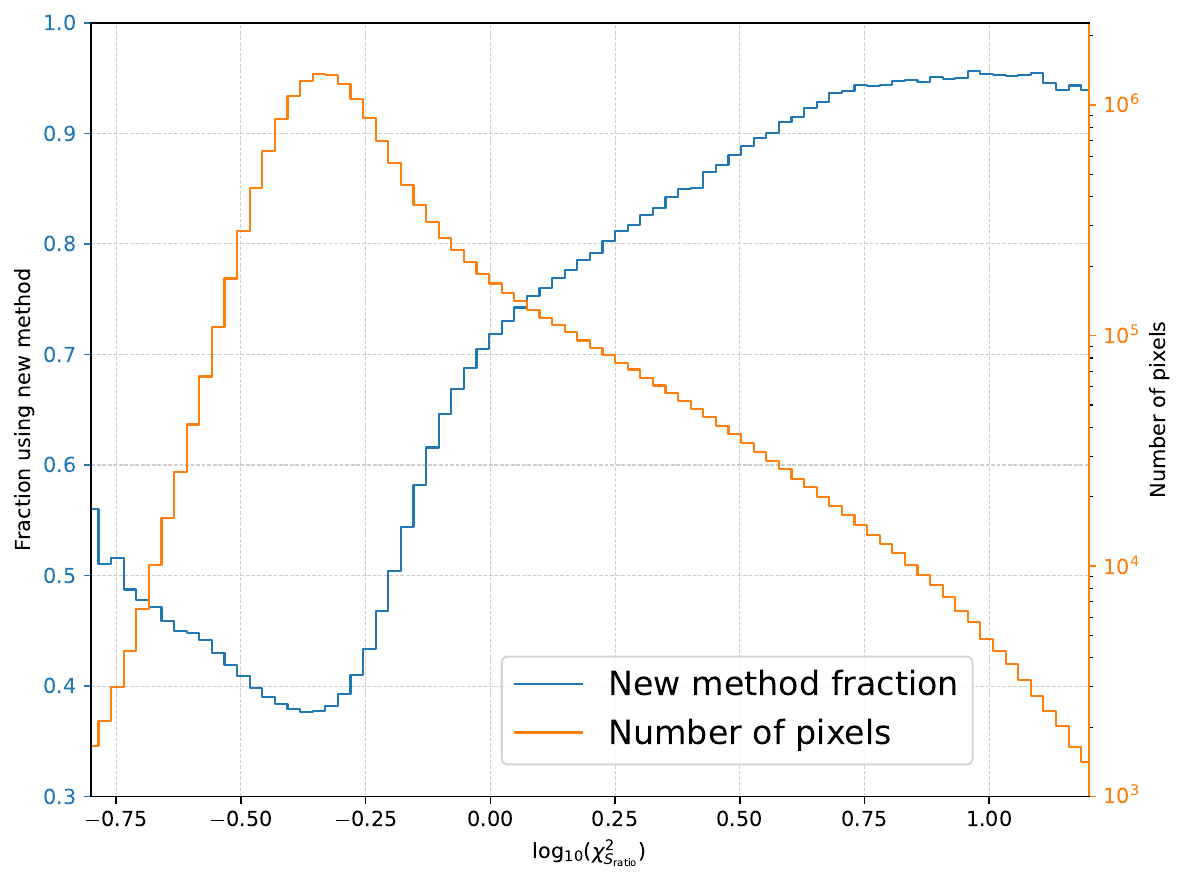}
    \caption{
    The blue histogram shows the fraction of pixels for which the new method was selected in each bin of $\log_{10}(\chi^2_{S_{\rm ratio}})$, while the orange histogram indicates the number of pixels in that bin.
    }
    \label{fig:NLC_ratio}
\end{figure}

Finally, we examine how the two NLC methods contribute to the combined approach by introducing a diagnostic parameter, $\chi^2_{S_{\rm ratio}}$, which measures the validity of the constant $S_{\rm ratio}$ assumption underlying the conventional model. From Eqs.~\eqref{eq:8} and \eqref{eq:S_ratio}, we have $S_{\rm ratio} = (A_2 + A_{\rm def}) / (A_1 + A_{\rm def})$. 
The assumption of a constant $S_{\rm ratio}$ therefore holds when $A_{\rm def}$ is constant and independent of the input illumination, or when its contribution is negligible compared with $A_1$ and $A_2$. We define $\chi^2_{S_{\rm ratio}}$ as the \rh{reduced} chi-squared statistic for fitting $S_{\rm ratio}$ with a constant model, where small values ($\chi^2_{S_{\rm ratio}}\!\lesssim\!1$) indicate that the conventional assumption remains valid. Figure~\ref{fig:NLC_ratio} shows that the fraction of pixels for which the new method is selected increases with increasing $\chi^2_{S_{\rm ratio}}$ beyond $\log_{10}\chi^2_{S_{\rm ratio}}\!\gtrsim\!-0.3$, demonstrating that the new method becomes dominant where the conventional assumptions fail, that is, in pixels affected by non-negligible or nonlinear defect currents. This result confirms that the combined approach adaptively selects the appropriate model on a pixel-by-pixel basis.

\section{Discussion and Conclusion \label{sec:conclusion}}

\begin{figure}[t!]
    \centering
    \includegraphics[width=\columnwidth]{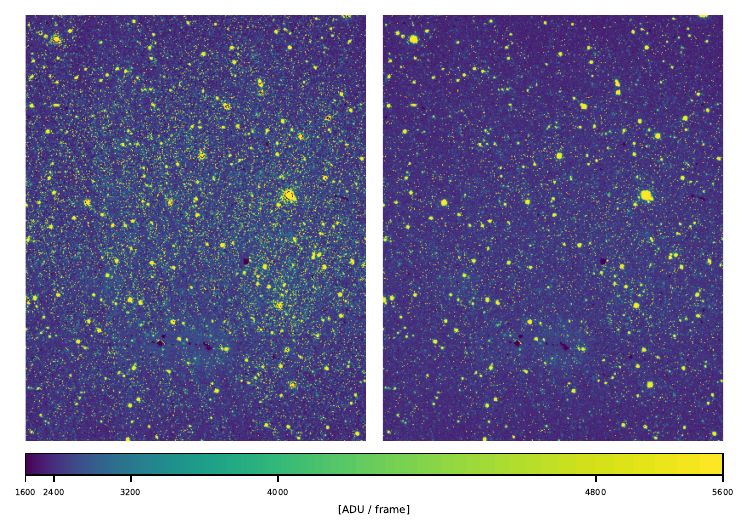}
    \caption{
    {\bf Left}: \rh{The same on-sky image shown in the \rev{middle} panel of Figure\ref{fig:comp_nlc},  processed with} superbias subtraction, reference pixel correction, conventional NLC, and ramp fitting, and dark subtraction, shows a cross-hatched pattern in certain regions. \gm{This pattern seems to arise from nonlinear defect currents in the corresponding pixels that are not accounted for in CNLC.} {\bf Right}: The \rh{on-sky} image processed with reference pixel correction, our proposed NLC (combined) method, and ramp fitting. The systematic pattern is visually mitigated, demonstrating the effectiveness of the new correction. \rev{The left and right panels share the same color scales as those in Figure~\ref{fig:comp_nlc}, with a stretch applied using a method similar to the DS9 {\it zscale} + {\it sinh} scaling.}
    }
    \label{fig:comp_nlc_result}
\end{figure}

We have developed and evaluated a new NLC method that extends the classical framework by explicitly modeling nonlinear defect currents that are not captured by conventional approaches. This formulation substantially improves correction accuracy for pixels affected by the defect currents, while maintaining consistency with the conventional model for well-behaved pixels. \rev{However, the enhanced flexibility of the new method comes at the cost of increased computational complexity. In particular, the calculation of $R_{\mathrm{meas}}$ requires approximations and multiple arithmetic operations, which can amplify its uncertainty compared to the conventional method. Hence, the conventional correction method would be more stable for pixels with negligible defect currents.} %The enhanced flexibility, however, comes at the cost of higher computational complexity and greater sensitivity to noise, making the conventional method more stable for pixels with negligible defect currents. 
A key result of this study is that optimal correction performance is achieved by combining both methods on a pixel-by-pixel basis. In the H4RG-10 detectors on the PRIME-Cam (chip 4), we found that $\sim47$ \% \rh{of the pixels for which the nonlinearity correction can be validly computed} were better corrected by the new model, confirming that the defect-current-related nonlinearity is a substantial effect under elevated-temperature operation. The combined framework thus provides a practical path to maintain uniform linearity \gm{correction} across the full array.

Figure~\ref{fig:comp_nlc_result} shows the suppression of the cross-hatch pattern in the processed images, demonstrating that the new correction effectively reduces spatially correlated artifacts. \gm{The origin of crosshatch is theorized to originate from the morphological variations of the HgCdTe surface along the 3 crystal axes due to growth conditions (\cite{Martinka2001}). These sub-optimal variations in the crystal structure of the detector material can cause elevated dark current along the 3 crystal axes. The observed suppression of spatial artifacts from elevated dark in the crosshatch pattern \rev{(as seen in the right panel of Figure~\ref{fig:comp_nlc})} with the new NLC correction indicates that the behavior of elevated dark current in these crosshatch pixels is well matched to our proposed model.}

Modern infrared detectors are designed to operate under nominal thermal and electronic conditions. However, post-launch environments can depart from these expectations due to cooling degradation, radiation damage, or long-term aging. Under such circumstances, hardware mitigation may be infeasible, and software-based approaches become essential. The method developed here demonstrates that accurate signal linearization can still be achieved by explicitly modeling pixel-level nonlinearities, including defect-current effects. This capability provides a practical contingency calibration strategy for maintaining detector linearity when ideal operating conditions cannot be ensured. More broadly, the results highlight that detector non-idealities, traditionally addressed through design, can be effectively corrected post facto through data-driven modeling, offering a robust means to preserve scientific performance in future space missions.

\section*{Acknowledgments}
We appreciate Naoki Koshimoto for fruitful discussions related to this work. We also acknowledge the contribution of \rh{students of The University of Osaka} for their support in data acquisition. 
The PRIME project is supported by JSPS KAKENHI Grant Nos.\ JP16H06287, JP22H00153, JP25H00668, JP19KK0082, JP20H04754, JP24H01811, and by the JSPS Core-to-Core Program (Grant No.\ JPJSCCA20210003). We acknowledge financial support by the Astrobiology Center.

\software{ 
\texttt{numpy} \citep{harris2020array}; \texttt{scipy} \citep{2020SciPy-NMeth}
}

\appendix

\section{Numerical Implementation with Discrete Data} \label{ap:A}

This section describes the numerical implementation of the procedures introduced in Section~\ref{subsec:new_method}, adapted for discretely sampled data. In practice, the measured signals $\{S_\mathrm{meas}\}$ are recorded at discrete time intervals, and the sampling pattern differs among the datasets $\{S_{\mathrm{meas},A_j}\}$ acquired under different illumination levels $A_j$. These conditions introduce several numerical challenges: (i) estimating the temporal derivative $\dot{S}_{\mathrm{meas}}$ from discretely sampled frames, (ii) evaluating the differences and ratios of $\{\dot{S}_{\mathrm{meas},A_j}\}$ at matched values of $\{S_{\mathrm{meas}}\}$, and (iii) performing numerical integrations, such as those required in Eq.~\eqref{eq:13}. We also describe additional preprocessing steps applied specifically to the dark-exposure dataset $\{S_{\mathrm{meas},A3}\}$.

\subsection{Finite-Difference Evaluation of $\dot{S}_{\mathrm{meas}}$}

To compute $\dot{S}_{\mathrm{meas},A_j}$ from discrete frame data, we evaluate the derivative at the midpoints between adjacent frames.  
The midpoint signal level is defined as
\begin{align}
    S'_{\mathrm{meas},A_j,i} = \frac{1}{2}\left(S_{\mathrm{meas},A_j,i+1} + S_{\mathrm{meas},A_j,i}\right), \label{eq:a.1}
\end{align}
and the corresponding time derivative is estimated by a finite difference:
\begin{align}
    \dot{S}_{\mathrm{meas},A_j}(S'_{\mathrm{meas},A_j,i}) =
    \frac{S_{\mathrm{meas},A_j,i+1} - S_{\mathrm{meas},A_j,i}}{\Delta t_i},\label{eq:a.2}
\end{align}
where $\Delta t_i = t_{i+1} - t_i$. This midpoint formulation provides a discrete analog of Eq.~\eqref{eq:14}, expressed as
\begin{align}
    S_{\mathrm{meas},A_j,i}
    = S_{\mathrm{meas},A_j,0}
    + \sum_{k=0}^{i-1} \dot{S}_{\mathrm{meas},A_j}(S'_{\mathrm{meas},A_j,k})\,\Delta t_k. \label{eq:a.3}
\end{align}
Thus, the accumulated signal in frame $i$ is obtained as the discrete integral of the temporal derivative evaluated at the midpoint between adjacent frames, ensuring consistency with the continuous formulation.

\subsection{Discrete Evaluation of $\Delta\dot{S}_{\mathrm{meas}}$ and $S_{\mathrm{ratio}}$ at Matched Signal Levels}

Because the interpolated signal grids $S^\prime_{\mathrm{meas},A_1}$ and $S^\prime_{\mathrm{meas},A_2}$ do not generally coincide, the time derivatives $\dot{S}_{\mathrm{meas},A_j}$ must be compared at the nearest available signal levels. For each $S^\prime_{\mathrm{meas},A_1,i}$, we identify the closest point in the other dataset as
\begin{equation}
    \ell = \arg\min_{\ell}\left|S^\prime_{\mathrm{meas},A_2,\ell} - S^\prime_{\mathrm{meas},A_1,i}\right|. \notag
\end{equation}

For brevity, let $\dot{S}_{j,i} \equiv \dot{S}_{\mathrm{meas},A_j}(S'_{\mathrm{meas},A_j,i})$. 
Then, the differential and ratio quantities are given by
\begin{align}
    \Delta\dot{S}_{A_1,A_2,i} &\simeq 
    \dot{S}_{1,i} - \dot{S}_{2,\ell}, \label{eq:a.4}\\
    S_{\mathrm{ratio},i} &\simeq 
    \frac{\dot{S}_{2,\ell}}{\dot{S}_{1,i}}. \label{eq:a.5}
\end{align}

This procedure enables consistent estimation of $\Delta\dot{S}_{\mathrm{meas}}$ and $S_{\mathrm{ratio}}$ even when the sampling grids of the two datasets are not perfectly aligned in $S_{\mathrm{meas}}$.

\subsection{Numerical Estimation of $A_{\mathrm{def}}$}

As described in Eq.~\eqref{eq:Adef}, the defect-current term $A_{\mathrm{def}}$ is derived from the dark (defect-dominated) exposure sequence $\{S_{\mathrm{meas},A_3}\}$ using the fitted nonlinearity function $R_{\mathrm{fit}}(S_{\mathrm{meas}})$. 
The measured signal is first corrected for classical nonlinearity as
\begin{equation}
    S_{\mathrm{corr},A_3,i}
    = S_{\mathrm{meas},A_3,i}\cdot R_{\mathrm{fit}}\!\left(S_{\mathrm{meas},A_3,i}\right),
\end{equation}
and the instantaneous defect current is then obtained by a finite-difference derivative,
\begin{equation}
    A_{\mathrm{def},i}\!\left(S'_{\mathrm{meas},A_3,i}\right)
    = \frac{S_{\mathrm{corr},A_3,i+1} - S_{\mathrm{corr},A_3,i}}{\Delta t_i}. \label{eq:a.7}
\end{equation}
This procedure provides a stable and self-contained estimate of $A_{\mathrm{def}}$ without requiring interpolation across illumination levels or reference to other datasets. 
Because the correction via $R_{\mathrm{fit}}$ removes the classical nonlinearity, the resulting $A_{\mathrm{def}}$ effectively represents the illumination-independent component of the defect current. 
The discrete estimates $A_{\mathrm{def},i}$ are finally modeled as a smooth function of $S_{\mathrm{meas}}$ to obtain $A_{\mathrm{def,fit}}(S_{\mathrm{meas}})$, 
which is then used in Eq.~\eqref{eq:18} to reconstruct the linearized signal $S_{\mathrm{lin}}$. Additionally, $S_{\rm def,lim}$ is chosen as the maximum $S_{\mathrm{meas},A_3,i}$ below which $A_{\rm def,fit}$ is always non-negative.

\subsection{Numerical Integration Procedures}

Using the relations derived above, the quantity $S_{\mathrm{lin+def}}$, originally defined in Eq.~\eqref{eq:13}, can be discretized as 
\begin{align}
    &S_{\mathrm{lin+def},A_1,i} - S_{\mathrm{lin+def},A_1,0} \notag\\
    &=  (A_1 - A_2) \sum_{k=0}^{i-1} \left[1 - S_{\mathrm{ratio}}\left(S^{\prime}_{\mathrm{meas},A_1,k}\right)\right]^{-1} \Delta t_k. \label{eq:a.8}
\end{align}
The ratio $R_{\mathrm{meas}}$ is then obtained as \footnote{In practice, only data with $i \ge 1$ are used, and the initial offset term
$(S_{\mathrm{lin+def},A_1,1}-S_{\mathrm{meas},A_1,1})/S_{\mathrm{meas},A_1,i}$
is neglected, since $S_{\mathrm{lin+def},A_1,1}-S_{\mathrm{meas},A_1,1} \ll S_{\mathrm{meas},A_1,i}$ for all $i$.
}
\begin{align}
    &R_{\mathrm{meas},i} = \frac{S_{\mathrm{lin+def},A_1,i}}{S_{\mathrm{meas},A_1,i}} \notag \\
    &= \frac{
        S_{\mathrm{lin+def},A_1,0} 
        + \sum\limits_{k=0}^{i-1}\left[
        \frac{A_1 - A_2}{
            1 - S_{\mathrm{ratio}}\left(S^{\prime}_{\mathrm{meas},A_1,k}\right)
        } \Delta t_k \right]
    }{
        S_{\mathrm{meas},A_1,0}
        + \sum\limits_{k=0}^{i-1} \left[
        \dot{S}_{\mathrm{meas},A_1}\left(S^{\prime}_{\mathrm{meas},A_1,k}\right) \Delta t_k \right]
    } \notag \\
    &= 1 
    + \frac{
        S_{\mathrm{lin+def},A_1,0}
        - S_{\mathrm{meas},A_1,0}
    }{
        S_{\mathrm{meas},A_1,i}
    } \notag \\
    &\quad + \frac{
        \sum\limits_{k=0}^{i-1} 
        \left[
            \frac{A_1 - A_2}{
                1 - S_{\mathrm{ratio}}\left(S^{\prime}_{\mathrm{meas},A_1,k}\right)
            }
            - \dot{S}_{\mathrm{meas},A_1}\left(S^{\prime}_{\mathrm{meas},A_1,k}\right)
        \right] \Delta t_k
    }{
        S_{\mathrm{meas},A_1,i}.
    } \label{eq:a.9}
\end{align}

We next describe the numerical evaluation of the integral of $A_{\mathrm{def,fit}}$ in Eq.~\eqref{eq:18}. When the temporal sampling is dense and $A_{\mathrm{def,fit}}$ varies nearly linearly between two adjacent frames, the integral can be approximated using the trapezoidal rule:
\begin{align}
    &\int_{t_{i-1}}^{t_i} A_{\mathrm{def,fit}}\!\left(S_{\mathrm{meas}}(t)\right) dt \notag\\
    &\simeq \frac{\Delta t_{i-1}}{2}\left[A_{\mathrm{def,fit}}\!\left(S_{\mathrm{meas},i}\right) + A_{\mathrm{def,fit}}\!\left(S_{\mathrm{meas},i-1}\right)\right]. \label{eq:a.10}
\end{align}
Alternatively, because $A_{\mathrm{def,fit}}$ is defined explicitly as a function of $S_{\mathrm{meas}}$, a change of variables allows the time integral to be rewritten as
\begin{align}
&\int_{t_{i-1}}^{t_i} 
   A_{\mathrm{def,fit}}\!\left(S_{\mathrm{meas}}(t)\right) dt \notag\\
&= \int_{S_{\mathrm{meas},i-1}}^{S_{\mathrm{meas},i}}
   A_{\mathrm{def,fit}}\!\left(S_{\mathrm{meas}}\right)
   \left[\dot{S}_{\mathrm{meas}}\!\left(S_{\mathrm{meas}}\right)\right]^{-1}
   dS_{\mathrm{meas}}. \label{eq:a.11}
\end{align}
Since $\dot{S}_{\mathrm{meas}}$ is available only at discrete points, it is approximated as constant within each interval, yielding
\begin{align}
&\int_{S_{\mathrm{meas},i-1}}^{S_{\mathrm{meas},i}}
   A_{\mathrm{def,fit}}\!\left(S_{\mathrm{meas}}\right)
   \left[\dot{S}_{\mathrm{meas}}\!\left(S_{\mathrm{meas}}\right)\right]^{-1}
   dS_{\mathrm{meas}} \notag\\
&\simeq
   \frac{\Delta t_{i-1}}{S_{\mathrm{meas},i}-S_{\mathrm{meas},i-1}}
   \int_{S_{\mathrm{meas},i-1}}^{S_{\mathrm{meas},i}}
   A_{\mathrm{def,fit}}\!\left(S_{\mathrm{meas}}\right)\,dS_{\mathrm{meas}}. \label{eq:a.12}
\end{align}
This formulation provides an equivalent estimate of the defect-current contribution to the total signal increment between frames. \rh{It can be particularly useful when the temporal spacing or nonlinearity invalidates the simple trapezoidal approximation.} In this study, the linearity was found to be nearly identical for both integration methods in Eqs.~\eqref{eq:a.11} and ~\eqref{eq:a.12}, so the trapezoidal integration is employed.

\section{Uncertainty Estimation\label{ap:B}}

We estimate the uncertainties of the measured signal $S_{\mathrm{meas}}$ and its time derivative $\dot{S}_{\mathrm{meas}}$, assuming Poisson-dominated shot noise and temporally uncorrelated readout noise.

\subsection{Frame-level Signal Uncertainty}
The Analog-to-Digital Unit (ADU), the number of photoelectrons (or accumulated charge) $Q$ [$e^{-}$], and the system gain $g$ [$e^{-}$/ADU] are related by $\mathrm{ADU} = Q/g$.
Assuming that the photoelectron generation follows Poisson statistics ($\sigma_Q^2 = Q$), the variance in digital units is
\begin{equation}
\sigma_{\mathrm{DN}}^2 = \frac{\sigma_Q^2}{g^2} = \frac{Q}{g^2} = \frac{\mathrm{ADU}}{g}.
\end{equation}
This relation provides the signal uncertainty in the photon-limited regime.

Let $d_i$ be the inter-frame increment between frames $i-1$ and $i$, and $\epsilon_i$ the random readout offset in frame $i$.  
Then the accumulated signal is
\begin{equation}
S_{\mathrm{meas},i} = \sum_{k=0}^{i} d_k + \epsilon_i .
\end{equation}
Since each increment $d_k$ arises from a Poisson process, its variance is $\sigma^2_{d_k} = d_k / g$.
Summing the independent terms gives
\begin{align}
\sigma^2_{S_{\mathrm{meas},i}} &= \sum_{k=0}^{i} \sigma^2_{d_k} + \sigma^2_{\epsilon}
\simeq \frac{S_{\mathrm{meas},i}}{g} + \sigma^2_{\epsilon},
\end{align}
where we adopt $g = 1.8\,[e^{-}/\mathrm{ADU}]$ and $\sigma_{\epsilon} = 5.0\,[\mathrm{ADU}]$ as representative values.

\subsection{Slope-level Uncertainty}

The frame-to-frame slope at $S^{\prime}_{\mathrm{meas},i}$ is evaluated via finite differences:
\begin{equation}
\dot{S}_{\mathrm{meas}}(S^{\prime}_{\mathrm{meas},i}) = \frac{1}{\Delta t}\left(d_{i+1} + \epsilon_{i+1} - \epsilon_i\right).
\end{equation}
The corresponding variance is
\begin{align}
\sigma^2_{\dot{S}_{\mathrm{meas}}(S^{\prime}_{\mathrm{meas},i})}
\simeq \frac{1}{(\Delta t)^2}\left(\frac{S_{\mathrm{meas},i+1} - S_{\mathrm{meas},i}}{g} + 2\sigma^2_{\epsilon}\right).
\end{align}
This incremental formulation naturally accounts for the shot-noise covariance between successive frames.  
We neglect temporal correlations in $\epsilon_i$ and any small covariance between frames when fitting the ramp samples. If higher precision is required, the full covariance matrix should be incorporated in a generalized least-squares approach \citep[e.g.,][]{Brandt2024}.

Finally, we note that this treatment ignores possible gain variations associated with detector nonlinearity. Nevertheless, the derived variances provide practical estimates for the relative uncertainties in $S_{\mathrm{meas}}$ and $\dot{S}_{\mathrm{meas}}$, which are subsequently propagated to quantities such as $S_{\mathrm{ratio}}$ and $R_{\mathrm{meas}}$.

\clearpage
\bibliographystyle{aasjournal}
\bibliography{reference}

\end{document}